\documentstyle[aps,twocolumn,psfig]{revtex}

\newcommand{\ms}	{$M_s$}
\newcommand{\mso}	{$M_s^0$}
\newcommand{\np}	{nti-phase domain} 
\newcommand{\lscoy}	{$\rm La_{2-y} Sr_y Cu O_4$}
\newcommand{\la}       {$^{139}$La}
\newcommand{\lc}       {lanthanum cuprate}
\newcommand{\lilco}    {$\rm La_2 Cu_{1-x} Li_{x} O_{4}$}
\newcommand{\lilcox}   {$\rm La_2 Cu_{1-x} Li_{x} O_{4}$}
\newcommand{\tn}       {$T_N$}
\newcommand{\etal}     {{\it et al}}
\newcommand{\cuot}     {CuO$_2$}
\newcommand{\lsco}     {$\rm La_{2-x} Sr_x Cu O_4$}
\newcommand{\lnsco}    {$\rm La_{1.48} Nd_{0.4} Sr_{0.12} Cu O_4$}
\newcommand{\beq}      {\begin{equation}}
\newcommand{\eeq}      {\end{equation}}

\title{Mobile Anti-phase Domains in Lightly Doped Lanthanum Cuprate}
\author{P. C. Hammel, B. J. Suh and J. L. Sarrao}
\address{Condensed Matter and Thermal Physics, 
Los Alamos National Laboratory, Los Alamos, NM 87545}
\author{Z. Fisk}
\address{National High Magnetic Field Laboratory, Florida State University, Tallahassee, FL 32306}
\author{\small(Received:  \quad May 21, 1998)}

\address{
\parbox{14cm}{\bigskip\rm\small
Light hole doping of \lc\ strongly suppresses the onset of 
antiferromagnetic (AF) order.  
Surprisingly, it simultaneously suppresses the extrapolated zero temperature
sub-lattice magnetization.
\la\ NQR results in lightly doped \lilcox\ have demonstrated
that these effects are independent of the details of the mobility 
of the added holes.
We propose a model in which doped holes phase separate into 
charged domain walls that surround ``anti-phase'' domains.
These domains are mobile down to $\sim 30\,$K where they either become 
pinned to the lattice or evaporate as their constituent 
holes become pinned to dopant impurities.
}}

\begin{document}
\maketitle


\section{Introduction}
A fundamental issue in the normal state 
of the superconducting cuprates is the behavior 
of holes doped into a two-dimensional lattice of spins with 
strong antiferromagnetic (AF) interactions.  
Even for lightly doped, single layer \lc\
many important issues remain poorly understood. 
Long-range antiferromagnetic order occurs at \(T_N > 300\) K in undoped \lc, but  
\tn\ is rapidly suppressed by the addition of a small density, $p$ of holes per Cu.  
This rapid suppression 
is clearly related to the disruptive effects of mobile holes: 
$p \lesssim 3\%$ is sufficient to suppress \tn\ to zero, 
while $\sim 30\%$ isovalent substitution of Zn or Mg for Cu is required\cite{swc:zn} 
to produce the same effect.  
A range of studies\cite{sarrao:prb96} including \la\ NQR 
measurements\cite{h:lilcolite} 
in lightly doped \lilcox\ have demonstrated
that the suppression of \tn, 
and in fact, all the magnetic properties of lightly doped \lc\ 
are essentially {\it invariant} without regard for the means of hole doping 
and consequent variations in hole mobility. 

It is unlikely that a collection of individual holes can lead to 
magnetic behavior that is entirely independent of compositional 
variation that leads to substantial variations in resistivity (at constant doping).  
We argue, instead, that this is strong evidence that holes form collective structures. 
An important and well documented aspect of doped cuprates is their tendency toward  
inhomogeneous charge distribution\cite{phasesep93}.  
Segregation of doped holes into charged stripes separating hole-free domains 
has been 
predicted\cite{str:zg,str:rice89,str:schulz89,ek:prl90:phsep,str:rice94,str:machida,str:zaanen:rev97} 
and recently observed directly in \lc\cite{tr:stripe:nature}.  
It was proposed earlier that phase segregation of holes 
could be responsible for the 
unusual magnetic properties of lightly 
Sr-doped \lc\cite{cho:prl93,chou:prl93,ames:b:prb95}. 
We make a related proposal that holes form charged, domain walls which form closed loops
with the important differences that these walls form 
anti-phase domain walls (so the the phase of the AF order inside these domains is reversed) 
and that the walls and hence the enclosed domains are mobile, 
and the charged walls have the density of 1 hole per 2 Cu sites in agreement 
with neutron scattering results\cite{tr:stripe:nature}. 
The anti-phase character means that mobile (above 30 K) domains 
will suppress the time-averaged static moment thus suppressing \ms\ as well as \tn.  
These domain structures will have contrasting interactions with in-plane 
{\it vs.\ }out-of-plane dopants 
({\it e.g.,\/} stronger scattering by in-plane impurities) 
which explain the different transport behaviors, 
while the universal magnetic properties can be understood 
as long as the domains are sufficiently mobile that they move across a given site 
rapidly compared to a measurement time.

\section{Lightly doped lanthanum cuprate}\label{sec:lc}
A systematic study of the temperature 
$T$ and doping dependence of the static susceptibility in lightly doped \lc\ 
by Cho \etal. \cite{cho:prl93} provided evidence that the added 
holes are inhomogeneously distributed. 
The development of long-range antiferromagnetic order is signaled by a peak 
in the static susceptibility; 
they showed the rapid increase in the width of this peak with increasing hole density 
could be understood as arising from finite-size effects.
They proposed that doped holes form hole-rich domain walls
that bound hole-free domains, 
thus cutting off spin interactions across the boundary and 
truncating the growth of the spin-spin 
correlation length with decreasing temperature above \tn.
They deduced the doping 
dependence of the dimension $\cal L$ of the hole free-regions, 
and found \( {\cal L} \simeq (0.02/p)^2 \); this suggests that the density of holes 
within the boundary stripe is very low, \( \sim 1\) hole per 5 or 10 Cu sites. 

\la\ NQR measurements in lightly Sr-doped \lc\ by Chou \etal. 
provided a detailed picture of the $T$ and $p$-dependence of its 
magnetic properties\cite{chou:prl93}.
They found that for 30 K \(\lesssim  T < T_N \) the 
sublattice 
\begin{figure}
\psfig{file=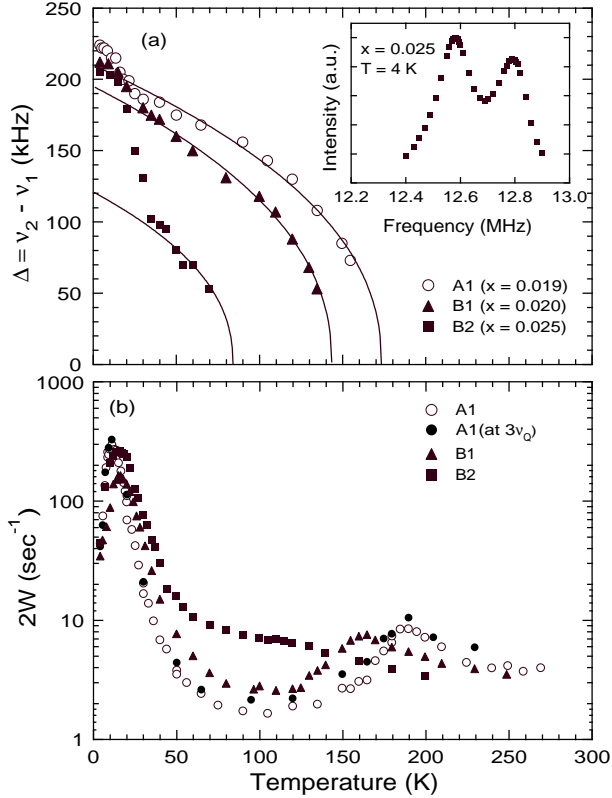,width=3.5in}
\caption{$^{139}$La NQR in La$_2$Cu$_{1-x}$Li$_x$O$_4$:
(a)~$\protect\Delta \protect\equiv \protect\nu _1 - \protect\nu _2$ vs. $T$.  Solid
curves are fits to the critical behavior 
$\Delta(T)=\Delta_0(1-T/T_N)^{\beta}$.  
The inset shows the split $2 \protect\nu _Q $ transition at 4 K 
The magnitude of this splitting measures the component 
of the internal field due to the AF ordered Cu moments 
parallel to the EFG axis.
(b)~$2W$ vs. $T$ is shown.  
The very strong peak in the vicinity of 15 K is very similar to 
that seen in Sr-doped \lc\protect\cite{chou:prl93}.}
\label{fig:data}
\end{figure}
\noindent
magnetization \ms\ is strongly suppressed as $p$ increases. 
However below 30 K, \ms\ recovers to its \( p = 0 \) value.  
The low temperature spin dynamics are also unusual; 
the \la\ nuclear spin-lattice relaxation rate 
$2W (\equiv 1/T_1)$ has a strong peak  
at a doping dependent temperature in the vicinity of 10--15 K.
To explain these unusual features, they extended the finite size model 
of Cho \etal.\cite{cho:prl93}, and proposed 
that the suppression of \ms\ could be understood 
in the context of the restricted set of spin-wave modes 
accessible in the confined AF domains\cite{ames:b:prb95}.
The low temperature peak in $2W$ is clearly associated 
with freezing of Cu spin degrees of freedom;
they interpreted this in terms of freezing out of hole motion 
within the domain walls surrounding hole-free regions.

Adding holes by means of in-plane substitution of 
Li$^{1+}$ for Cu$^{2+}$ introduces impurities into the \cuot\ 
planes which strongly alter charge transport properties. 
We have used \la\ nuclear quadrupole resonance (NQR) measurements to  
microscopically examine the effects of 
doped holes on the AF spin 
correlations (Fig.\ \ref{fig:data})
in this case where the hole mobility is much reduced compared 
\begin{figure}
\parbox{2mm}{\rule{0mm}{25mm}}
\parbox{3in}{
\psfig{file=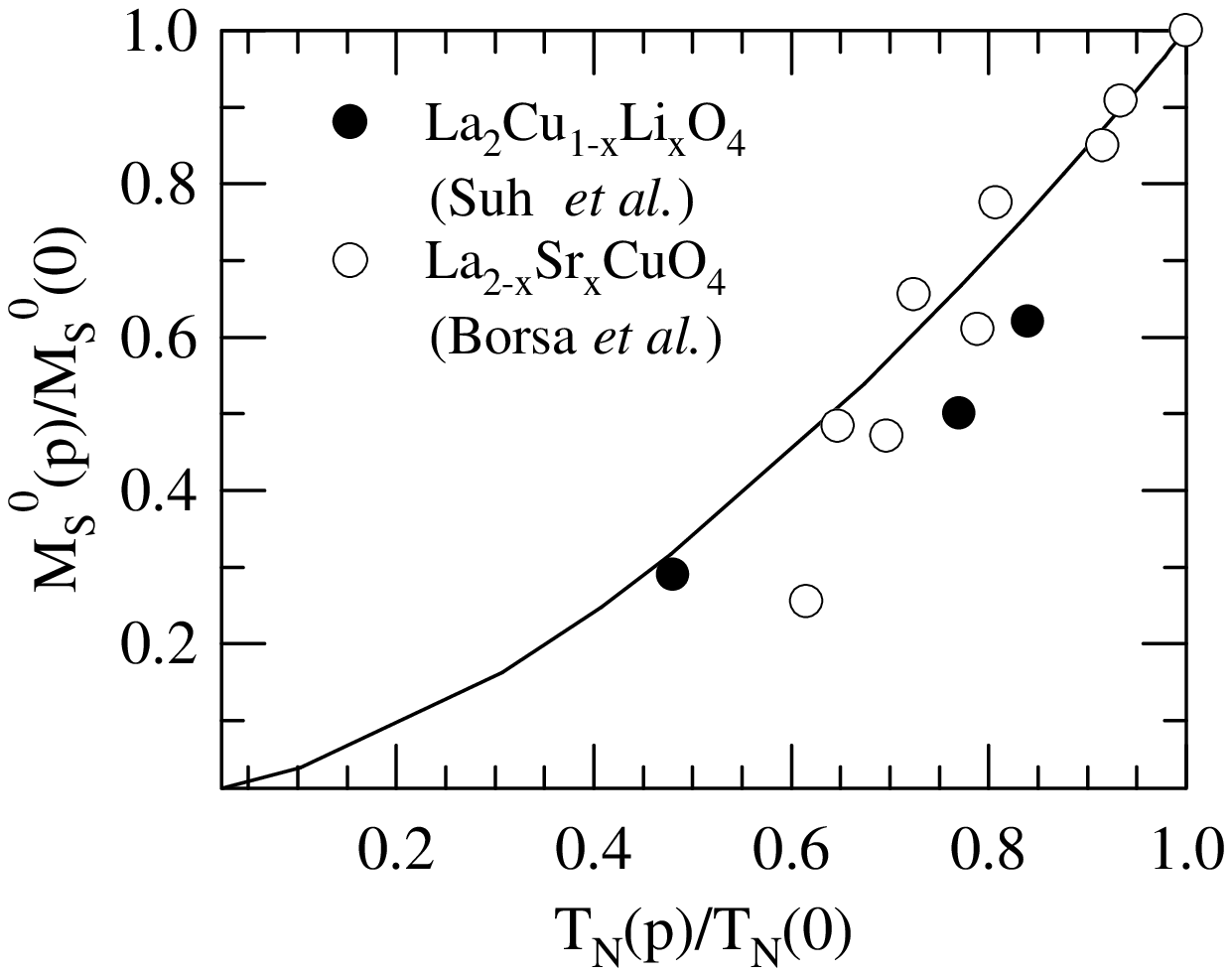,width=2.8in}}
\protect\vspace{3mm}\caption{The relationship between $ M_s^0(p) $ and 
of $ T_N(p) $  as both are suppressed by 
increasing doping $p$.  
$ M_s^0(p) $  is obtained as explained in the text from 
$^{139}$La NQR data for $ \protect\Delta \protect\nu $ 
such as is shown for the Li case in Fig.\ \protect\ref{fig:data}.  
The closed circles are from \lilco\protect\cite{h:lilcolite}, 
and the open circles are from \lsco\protect\cite{ames:b:prb95}.
The results are the same in both materials,
illustrating one aspect of the similarity of the magnetic properties, in spite of the 
much higher resistivity found in \lilco\ as a consequence of the in-plane
impurities. 
The solid line is due to Castro Neto and Hone\protect\cite{str:neto:prl96}.}
\label{fig:mstn}
\end{figure}
\noindent
to the Sr doping case\cite{h:lilcolite}.
Comparing \lscoy\ (LSCO) and \lilco\ (LCLO) at \( x = y = p = 0.025 \) 
one finds that the room temperature resistivity of LCLO\cite{sarrao:prb96,kastner} 
exceeds that of LSCO\cite{takagi} by over an order of magnitude.
Furthermore, unlike LSCO, the resistivity of LCLO always increases 
monotonically with decreasing temperature. 
With increasing doping the contrast becomes more dramatic 
as LSCO becomes metallic and superconducting while LCLO 
becomes ever more insulating with doping above \( p = 0.1 \).

In spite of this we find that the magnetic behavior 
of the two materials is essentially identical\cite{h:lilcolite}.  
In addition to the similarly strong suppression of \tn\ by doping\cite{sarrao:prb96}, 
we find that \ms\ is also suppressed, 
and the correspondence between the suppression 
of \ms\ and \tn\ by doping is identical to that observed in LSCO\cite{ames:b:prb95}.
In Fig.\ \ref{fig:mstn} \( M_x^0 (p)/M_x^0 (0) \) for both LCLO (Ref.\ \onlinecite{h:lilcolite}) 
and LSCO (Ref. \cite{ames:b:prb95}) is plotted against \( T_N (p)/T_N (0) \). 
Here \mso\ is the value of \ms\ obtained by extrapolating  
the \(M_s(T) \) data for $T > 30 \, $K to $T=0$ 
i.e., the \(T=0\) value of the solid lines shown in Fig.\ \ref{fig:data}(a).  
The solid line through the data is due to a theory of Castro Neto and 
Hone\cite{str:neto:prl96};
see also van Duin and Zaanen\cite{str:vz:prl97}.
The strong peak in $2W$ occurs at the 
same temperature and exhibits the same binding energy (as extracted 
from the $T$-dependence on the high temperature side of the peak)\cite{h:lilcolite}.
Finally, the temperature dependence of the low-energy dynamical 
susceptibility (obtained from measurements of $2W(T)$) 
exhibits the same finite-size effects\cite{h:lilcolite} 
as were observed in the static susceptibility by Cho \etal\cite{cho:prl93}. 

\section{Mobile anti-phase domains}
There is clear evidence for stripe formation in two-dimensional 
doped antiferromagnets.  
In La$_{2}$NiO$_{4}$ (isostructural to \lc) static 
stripes have been observed in several 
cases\cite{hayden:lsno,swc:lsno:prl93,tr:lno:prl94,tr:lno:prb96,swc:leeprl97}. 
The recent observation of similar elastic superlattice peaks 
in \lnsco\cite{tr:stripe:nature} 
demonstrates the existence of static charged stripes in the cuprates,
and supports the idea that stripes are universally present 
in \lc\cite{ek:prl95:badmetal,str:zaanenprb96}
but that they are observable as static only under special 
conditions which pin the stripes to the lattice\cite{tr:stripe:nature}. 
Similarities between elastic superlattice peaks associated with static stripes
and the incommensurate peaks observed in 
inelastic neutron studies of \lscoy\cite{swc:nslsco} have been noted,    
and these incommensurate peaks are being reconsidered as possible 
evidence for the presence of dynamic charged stripes in 
the cuprate\cite{tr:stripe:pc97}.
The density of holes in the charged domain walls depends on the material: 
in the nickelates it is 1 hole per stripe Ni site, 
and in the cuprate the density is 1/2 hole per stripe Cu site.
The neutron diffraction studies have demonstrated that 
spin-spin interactions are not cut off by the charged domain walls, 
rather interactions across them are strong: 
it is universally observed that they serve as anti-phase 
domain walls between the hole free regions they separate.
Thus the sign of the spin correlations is reversed upon 
crossing the domain wall.  

The formation of domain walls into loops as opposed to parallel stripes 
has been observed in Hartree-Fock calculations\cite{str:zg}.
Using density matrix renormalization group techniques to calculate the energy 
of a domain wall in the 2D $t$-$J$ model White and Scalapino have observed 
charged domain walls to form loops\cite{white:loops}.
They point out this is favorable at low doping in the 
case where the coupling between planes is significant.
Because the walls constitute anti-phase domain walls, 
the coupling between two planes is disrupted, in general, by domain walls.   

\begin{figure}
\parbox{15mm}{\rule{0mm}{25mm}}
\parbox{3in}{
\psfig{file=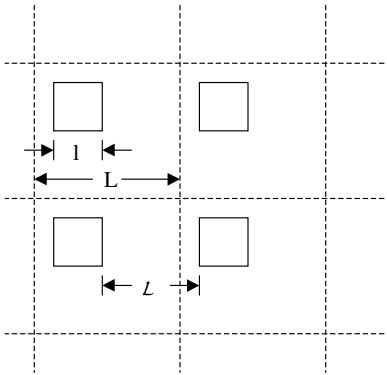,width=2in}}
\protect\vspace{5mm}\caption{Schematic diagram of the various lengths in the 
a\np\ model.  
On average, an area of dimension $L$ will contain a single a\np\ 
of dimension $l$.
The length $ {\protect\cal L} = L - l $ is the finite-size confinement length 
that should be compared to the results obtained by 
Cho \etal.\protect\cite{cho:prl93}.}
\label{fig:diag}
\end{figure}
\noindent
Hence, inter-plane coupling would favor domain walls forming  
closed loops so that most of each plane would be in the dominant AF phase.

In the event that these anti-phase domains are mobile, passage of such 
a domain over a given site will reverse the orientation of a 
particular ordered Cu moment. 
The splitting of the \la\ NQR line 
(shown in Fig.\ \ref{fig:data}) 
is proportional to local hyperfine field due
to the ordered moment on the neighboring Cu site.  
If this moment is time-varying, the splitting will be proportional 
to the time-averaged local moment.
In the absence of a\np s, the hyperfine field will be constant giving the 
value of \ms\ observed in undoped \lc. 
If the motion of the a\np s is rapid compared to the NQR measurement time, 
the net local hyperfine field will be proportional to the fraction of time
the moment is in the dominant AF phase minus the time it is in an a\np, 
and hence proportional to the area of the dominant phase minus the area 
of the a\np.
We can estimate the doping dependence of the size and spacing of the a\np s
from the known behavior of \mso\ in \lscoy\cite{ames:b:prb95}.  
If we define \( R(p) \equiv M_s^0(p) / M_s^0(0) \) the data\cite{ames:b:prb95}  
for $M_s^0(p)$ is well described by 
\begin{figure}
\parbox{2mm}{\rule{0mm}{25mm}}
\parbox{3in}{
\psfig{file=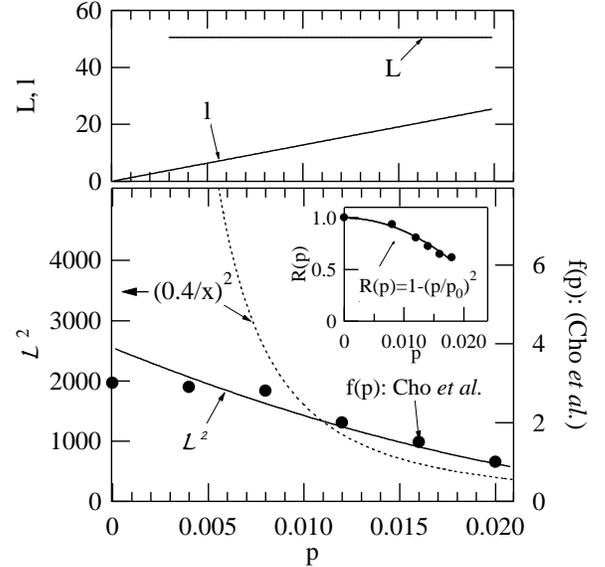,width=3in}}
\protect\vspace{3mm}\caption{Upper panel: 
The variation of the size $l$ of the a\np s 
required to explain the observed suppression of \mso\ by doping is shown 
along with the average size $L$ of the region which encompasses a single 
a\np. 
The inset to the lower panel shows the data of Borsa \etal . 
for $ R(p) \protect\equiv M_s^0(p) / M_s^0(0) $ \protect\cite{ames:b:prb95}
along with the parametrization of the $p$-dependence used to calculate 
the lengths shown here; $ p_0 = 0.028 $.
Lower panel:  The variation of ${\protect\cal L}^2$ with doping 
is plotted against the left-hand axis,  
and the results of Cho \etal. \protect\cite{cho:prl93} 
are plotted against the right-hand axis.  
The fit obtained by scaling the single parameter which sets the overall 
magnitude of $ f(p) $ obtained by Cho \etal . is very good.
Also shown, plotted against the left-hand axis, is the fit $ (0.4/x)^2 $ suggested 
by Cho \etal . above $ p = 0.01 $.}
\label{fig:ap}
\end{figure}
\noindent \beq R(p) = 1 - (p/p_0)^2 \label{eq:rp} \eeq
with \( p_0 = 0.028 \). 
For simplicity we assume that a (1,0) or (0,1) domain wall orientation is preferred, 
and so consider square domains. 

If a region of size $L$ contains, on average, one a\np\ of size $l$ 
(see Fig.\ \ref{fig:diag}, 
all lengths are in units of the lattice parameter), then 
\beq
R =  1- (2N_- / N) 
\label{RN} 
\eeq 
Here \( N = N_+ + N_- = L^2 \), where \( N_- = l^2 \) is the number of sites 
in the a\np\ and $N_+$ is the number of sites in the dominant AF phase. 
The number of holes in the region of size $L$ is $pL^2$; 
the domain wall which bounds the a\np\ contains 1 hole per 2 Cu sites, 
so \( 4l = 2pL^2 \).  From Eqns.\ \ref{eq:rp} and \ref{RN}
\beq L^2 = N = 2 (1-R) / p^2 = 2 / p_0^2 \eeq
and
\beq l = (1-R)/p = p/p_0^2 \eeq
hence \( L \simeq 50 \). 
The variation of $l$ with $p$ based on the experimentally 
determined variation of $R(p)$ is shown in Fig.\ \ref{fig:ap}(a). 
It should be noted that the behavior found here is particularly
simple as a consequence of the parametrization of $R(p)$ chosen (Eq.\ \ref{eq:rp});
this parametrization is not uniquely determined by the data. 

This simple model has several appealing features.
The model described in Section \ref{sec:lc} which relies on 
static domain walls implies a very low hole density in the wall
(\(\sim 0.1 \)--0.2 holes/Cu site) which must nonetheless maintain its 
integrity as a charged stripe and entirely cut off AF interactions
across the stripe.  
Our model posits a density of 0.5 holes per Cu site such as is observed
in neutron scattering and predicted by calculations\cite{white:loops}. 
The recovery of \ms\ below 30 K is straightforwardly understandable 
since once motion of the a\np s becomes slow compared to the 
NQR time scale ($\sim 0.1$--1 $\mu$sec) time averaging of the reversed 
spin directions will cease and the full ordered moment will be observed. 
This could arise either from pinning of the a\np\ to the lattice 
or evaporation of the domain walls due pinning of the constituent holes 
to the charged donor impurities;
in either case the coincidence of the recovery of \ms\ 
and the freezing of spin degrees of freedom 
evidenced by the low $T$ peak in $2W$ is naturally explained. 
The correspondence between suppression of \mso\ and \tn\ is natural 
in this case because interlayer coupling will be hampered wherever 
an a\np\ is present, thus impeding the development three-dimensional AF ordering.
See the discussion in Ref.\ \onlinecite{white:loops} in this regard. 

This model also explains the finite-size effects revealed by the susceptibility
analysis of Cho \etal.\cite{cho:prl93} if we consider that the appropriate length 
scale between domain walls is \( {\cal L} = (L - l) \).
The variation of ${\cal L}^2$ with $p$ is shown in Fig.\ \ref{fig:ap}(b) 
and compared with the variation of the square of the characterisitic 
length scale obtained by Cho \etal.\cite{cho:prl93} (scaled vertically to obtain the 
best agreement).
Finally we note from Fig.\ \ref{fig:ap}(a), that $L$ and $l$ converge with increasing 
$p$, and we expect that loops will cease to be stable when $L$ approaches $l$.
For the parametrization of $R(p)$ we have chosen, 
\( L = l \) when \(p = \sqrt{2} \, p_0 = 0.04 \), near the 
doping at which the the metal-insulator transition and 
spin-glass behavior are found. 
We speculate, then, that these are related to the 
transition in the configuration of the charged domain walls from 
loops to parallel stripes.

In conclusion, we have presented a model which explains the 
range of unusual magnetic phenomena observed in lightly doped \lc.
In particular, we can understand the insensitivity of magnetic properties 
to materials variations that substantially increase the resistivity.  
This indicates that mobile a\np s play a central role in determining 
the magnetic properties of lightly doped \lc. 
It may point to an explanation of the poorly understood ``spin-glass''
regime of the phase diagram in terms of a crossover in domain wall topology 
from loops to parallel stripes.  
More generally, it suggests that the development of stripe order may play a 
determining role in the phase diagram of the cuprates
(see e.g., Ref.\ \onlinecite{kivelson:liqxtal}). 
Rather than requiring mobile domain walls, 
superconductivity may more sensitively depend on the nature of 
the ordering of the walls into parallel stripes. 

\section{Acknowledgements}
We gratefully acknowledge stimulating conversations with 
J. Zaanen, 
who suggested the idea behind the model presented here. 
Work at Los Alamos performed under the auspices of the US Department of Energy.  
The NHMFL is supported by the NSF and the State of Florida through 
cooperative agreement DMR 95-27035.


\end{document}